**Interpretation of cosmological expansion effects on the quantum-classical transition**

C. L. Herzenberg

**Abstract**
*Recently, what appears to be a fundamental limit on an object's size that separates the quantum behavior characterizing small objects from the classical behavior characterizing large objects has been derived from the Hubble velocity spread in an extended object. This threshold is now examined further and interpreted in terms of diffusion processes in stochastic quantum mechanics. This limiting size that separates quantum behavior from classical behavior is shown to correspond approximately to an object's diffusion distance over the Hubble time.*

**1. Introduction**

Although cosmic expansion at very small distances has in the past usually been dismissed as entirely inconsequential, recent work has established that these extraordinarily small effects may in fact have a real and significant influence on our world. Calculations have suggested that the minute recessional velocities associated with regions encompassed by extended bodies may have a role in creating the distinction between classical and quantum behavior.[1-3] What appears to be a fundamental threshold size that separates the quantum behavior characterizing small objects from the classical behavior characterizing large objects has been derived based on the intrinsic Hubble velocity spread in an extended object. We now examine in more detail the basis for this threshold size in the context of several related phenomena. Similar calculations appear to suggest the Hubble time may also have a role in determining this threshold size. Since the stochastic interpretation of quantum mechanic involves diffusion processes, we look into the diffusion processes in stochastic quantum mechanics to provide interpretation of an underlying physical basis for such a threshold size.

**2. Background: The influence of the Hubble velocity on the quantum-classical transition**

Since the cosmological recessional velocity or Hubble velocity between two locations is proportional to the distance separating them with the proportionality constant being the Hubble constant $H_o$, the spread in Hubble recessional velocities $\Delta v$ along any direction over an object of characteristic size L would reach an approximate value:[1-3]

$$\Delta v \approx H_o L \qquad (1)$$

If the spread in recessional velocities associated with an extended object is taken as a measure of an uncertainty in velocity, an associated uncertainty in momentum can be calculated, and from that an associated uncertainty in position Δx can be calculated from the Heisenberg uncertainty principle as:[1-3]

$$\Delta x \geq h/(4\pi m H_o L) \qquad (2)$$

Here h is Planck's constant, m is the mass of the object, $H_o$ is the Hubble constant, and L is a length characterizing the size of the object. To examine the critical limiting case or threshold value of the minimum uncertainty in position which is comparable to the linear dimension of the object, we can set the value of the minimum uncertainty in position equal to the length of the object and obtain as an estimate for the critical length $L_{cr}$ of the object:

$$L_{cr} \approx [h/(4\pi m H_o)]^{1/2} \qquad (3)$$

This threshold or critical length separates the class of objects whose size is greater than their uncertainty in position from the class of objects whose size is smaller than their uncertainty in position. For a given mass, objects of size greater than approximately this critical length would be expected to behave in a classical manner, while objects of appreciably smaller sizes could exhibit quantum behavior as entire objects, unless brought into classicity by environmental sources of quantum decoherence.[1-3]

## 3. Related effects

After having considered the implications in this context of the Heisenberg uncertainty relationship coupling an uncertainty in location to an uncertainty in momentum as summarized above, let us next consider the implications of the Heisenberg uncertainty relation coupling an uncertainty in energy to an uncertainty in time, $\Delta E \, \Delta t \geq h/4\pi$.

We can obtain an estimate of the uncertainty in the non-relativistic kinetic energy $E = \frac{1}{2} mv^2$ for an object of characteristic length L, by introducing approximate values for the Hubble velocity and its spread from Eqn (1):

$$\Delta E = mv \, \Delta v \approx m H_o^2 L^2 \qquad (4)$$

If we evaluate the magnitude of this uncertainty in energy for an object of a size corresponding to the critical length $L_{cr}$ by inserting the value for $L_{cr}$ from Eqn. (3) into Eqn. (4), we find:

$$\Delta E \approx (h/4\pi) H_o \qquad (5)$$

and thus we find as an estimate of the associated time uncertainty:

$$\Delta t \geq 1/H_o \qquad (6)$$

This relationship seems to be telling us that a spread in time of approximately the Hubble time (the inverse of the Hubble constant), which is roughly the lifetime of the universe, is associated with the threshold separating classical behavior from quantum behavior. How can that be understood?

In order to attempt to interpret this result and also to gain further understanding of the nature of the critical or threshold size, it may be informative to examine how a quantity that corresponds to the threshold length separating quantum and classical behavior could originate in a rather different context entirely, in terms of diffusion in a stochastic interpretation of quantum mechanics.

## 5. Stochastic quantum mechanics

The central idea of the stochastic interpretation of quantum mechanics is the view that uncertainty in a quantum particle's observable properties, such as position and momentum, is not due to the particle's lack of a sharp trajectory, but rather to the particle's following a random trajectory that is not directly observable by us. Thus, it is usually assumed that at a basic level, a particle actually behaves as a particle, and that at any specific time it has a well-defined position and a unique velocity, though there cannot be determined in practice.[4,5] Thus, stochastic interpretations of quantum mechanics generally treat quantum mechanics as fundamentally a classical theory of inherently probabilistic or stochastic processes. One can only determine their statistical distributions, and accept or demonstrate that these distributions are correctly described by ordinary quantum mechanics.[4] This generally involves microphysical particles performing some kind of Brownian motion, with the motion of a quantum mechanical particle interpreted in terms of classical trajectories and stochastic forces; furthermore, in stochastic quantum mechanics, we are dealing with a situation in which particles diffuse by a stochastic process, even in empty space.[4,6,7] This stochastic motion of particles is regarded as responsible for the statistical nature of quantum mechanics.

## 6. Stochastic particle diffusion by random walk/Brownian motion

Over a period of time, a particle in stochastic motion can be expected to undergo a diffusion type of displacement similar to Brownian motion or a random walk. We will start with a very simple and visualizeable approach to Brownian motion that will suffice for present purposes to introduce the main results of interest in the present connection.

For a simple random walk, the net root-mean-square (rms) straight-line distance or displacement x between start and finish positions that takes place over a time t is given by the equation: [8-10]

$$x_r(t) = (d_o^2 t/\tau)^{1/2} \tag{7}$$

where $d_o$ is the step length for a single individual displacement and $\tau$ is the time interval for a single displacement.

One approach to stochastic theories of quantum mechanics is the interpretation of nonrelativistic quantum mechanics as a variety of relativistic theory of Brownian motion.[11,12] In this approach, quantum behavior is regarded as derived from a stochastic theory of microphysical particles performing a kind of Brownian motion at or close to the speed of light.[6]

Let's first examine a random walk in a relativistic approach to Brownian motion, in which a particle undergoes a zitterbewegung-type stochastic motion at very high speeds close to the speed of light.[12] Such a particle of mass m is engaged in Brownian motion characterized by a frequency comparable to that particle's de Broglie frequency of $mc^2/h$. Accordingly, the average time interval associated with an individual displacement would be approximately equal to a time interval of $h/mc^2$. As a result, the particle would undergo random motions of these average time durations and average step lengths accordingly given approximately by the value $c\tau$ and thus by the particle's Compton wavelength, $h/mc$.

Using Eqn. (7) and inserting values for an average time interval associated with an individual displacement and the length of an individual displacement, we find that, after a large number of steps in a random walk, the particle's rms position relative to its original position would be given approximately by:

$$x_r(t) \approx (ht/m)^{1/2} \tag{8}$$

Eqn. (8) provides an estimate of the distance away from the starting location that a quantum object undergoing a random walk in such relativistic Brownian motion may be expected to move over a time t.

A more general approach can be used for any type of diffusion in three dimensions in terms of the variance and an equivalent diffusion constant, rather than relying on a particular detailed model for the Brownian motion.[8-10]

In stochastic diffusion in three dimensions, the rms distance or square root of the variance would be given by:

$$x_d(t)^2 = 6Dt \tag{9}$$

where D is the diffusion constant.[8-10]

In stochastic interpretations of quantum mechanics it is found that linearized equations formally identical with the Schroedinger equation can be obtained if the diffusion coefficient is taken as equal to $h/4\pi m$.[6,7,13] Thus, we may expect in the more general case of stochastic diffusion that:

$$x_d(t) = (3ht/2\pi m)^{1/2} \tag{10}$$

It can be seen that Eqn. (8) and Eqn. (10) give roughly similar estimates for the distance that a quantum object may be expected to diffuse over a time t.

The root mean square distance reached in a random walk will depend on the square root of the time available, as specified in Eqn. (8). If we take this time to be the Hubble time (that is, the inverse of the Hubble constant), which is roughly the lifetime of the universe, we will find for the net rms distance reached in a random walk process over the Hubble time:

$$x_r(t_{Ho}) \approx (h/mH_o)^{1/2} \tag{11}$$

and similarly, we find for the rms diffusion distance for a quantum object over the Hubble time from Eqn. (10) to be:

$$x_d(t_{Ho}) = (3h/2\pi mH_o)^{1/2} \tag{12}$$

These results appear to tell us that over the Hubble time or the lifetime of the universe, a quantum particle or object would undergo stochastic diffusion over a spatial region with a size comparable to the distances specified in Eqn. (11) or Eqn. (12).

Comparison between Eqn. (11) or Eqn. (12) and Eqn. (3) shows that the estimate of the critical length estimate $L_{cr}$ associated with the quantum-classical transition and the estimates made for random walk or diffusion displacement over the Hubble time are quite similar quantities with the same functional dependence on all of the same parameters, differing in value by only small numerical factors.

Thus, it appears that over the lifetime of the universe, a quantum particle or object would undergo stochastic diffusion over a spatial region of size approximately equal to the critical length associated with the quantum-classical transition.

What can be the significance of the similarity of these quantities that originate in such dissimilar contexts, and how are we to interpret this?

**7. Diffusion distance and regions in which quantum behavior occurs**

This stochastic interpretation of quantum mechanics appears to be telling us that a quantum particle or object in the course of its Brownian motion diffuses a certain distance from its original location over an interval of time. The particle diffusion distance can be characterized by and will not extend greatly beyond the rms distance given approximately by Eqn. (8) or Eqn. (10). Thus, throughout this period of time, the particle's location and its diffusive motion will be confined to a region largely within such a distance from its original position.

Quantum mechanical behavior as described in the stochastic interpretation of quantum mechanics originates from the presence of the stochastic or Brownian motion of the particle. Further, since in stochastic models of quantum mechanics, it is the Brownian motion of the object that leads to its quantum mechanical behavior, we might surmise that an object whose Brownian motion extended throughout the entire universe would be expected to behave as a quantum object everywhere. On the other hand, if the stochastic motion of the particle in question were only taking place in certain regions of space, it would appear that the quantum mechanical behavior of the particle would be limited to those regions. Thus, a free particle or object whose stochastic Brownian motion has throughout all of its history or throughout all time only involved a small region of space, might be expected to exhibit quantum behavior only within that region of space.

**8. Further discussion**

These results suggest that the quantum behavior of any object may be limited to within a region of approximate size determined by the quantity $L_{cr}$, the critical length determined by its mass. What are the magnitudes of these critical lengths, and the associated regions of space of presumed possible quantum behavior, for various objects? If we insert numerical rest mass values into Eqn. (3), we find the critical length associated with an electron would be about 5000 km, that for a proton about 120 km. An object with a mass of a nanogram would have a critical length of about a millimeter; while an object with a mass of a microgram would have a critical length of about 0.15 mm, and an object with a mass of a gram would have a critical length of about 15 nanometers. Thus, elementary particles and very low-mass objects would seem to have relatively large associated regions of space in which they might exhibit quantum behavior. On the other hand, larger objects in a more familiar size range will have any quantum behavior (as an entire object) restricted to much smaller regions of space, smaller than the physical size of the object above the critical size. (As noted in preceding work, while these considerations introduce basic limits, actual quantum-classical transitions can be significantly affected by ordinary environmental decoherence effects.)[1-3]

These results also suggest implications for the behavior of matter during the history of the universe. If we consider what comparable results would be at earlier or later times in the history of the universe, it would appear that a Hubble constant appropriate to the time in question should be used in the formulae. As a consequence, all critical lengths for earlier times would be smaller than they are today, while all critical lengths at later times would be larger than they are at present. Accordingly, we might expect that classical behavior would have been somewhat more widespread in the early universe, while on the other hand quantum behavior would be expected to be somewhat more widespread at later times during the expansion of the universe.

## 9. Summary and conclusions

A fundamental limit on an object's size separates the quantum behavior typically characterizing small objects from the classical behavior typically characterizing large objects. This derivation based on the Hubble velocity spread in an extended object has been examined further, and related results involving energy and time uncertainty have been developed. An analysis of diffusion in stochastic quantum mechanics seems to indicate that, for a given mass, the limiting size that separates quantum behavior from classical behavior will correspond closely to the diffusion distance experienced by an object over the Hubble time or the age of the universe.

These results suggest that the quantum behavior of an object will be limited to and manifested within a region of size determined by the quantity $L_{cr}$, the critical length determined by the mass of the object.

Extending the foregoing analysis to other epochs or time periods in the age of the universe besides our own time, results suggest that we might expect that classical behavior of objects would have been more prevalent in the early universe, while a more widespread extension of quantum behavior might be expected at later times during the expansion of the universe.

**Acknowledgments**

I would like to thank Norman H. Redington for pointing out after examining my preceding study that a parallel application of the time-energy uncertainty principle would lead to the interesting result of a time uncertainty comparable to the age of the universe, and for posing the question of an interpretation of this result.

**References**

1. C. L. Herzenberg, "Becoming classical: A possible influence on the quantum-to-classical transition," arXiv.org e-Print archive, physics/0602161 (23 February, 2006).

2. C. L. Herzenberg, "A possible cosmological effect on the quantum-to-classical transition," arXiv.org e-Print archive, physics/0603136 (16 March, 2006)

3. C. L. Herzenberg, "Becoming classical: A possible influence on the quantum-classical transition," submitted for publication to Physics Essays (May 17, 2006)

4. R. Omnes, *The Interpretation of Quantum Mechanics* (Princeton University Press, Princeton NJ, 1994).

5. E. Santos, "The Search for Hidden Variables in Quantum Mechanics," in *Quantum Mechanics versus Local Realism,"* (ed. F. Selleri) (Plenum Press, New York, 1988).

6. M. Jammer, *The Philosophy of Quantum Mechanics: The Interpretations of Quantum Mechanics in Historical Perspective* (Wiley, New York, 1974).

7. E. Nelson, Phys. Rev. **150**, 1079 (1966).

8. H.C. Berg, *Random Walks in Biology*, expanded edition (Princeton University Press, Princeton NJ, 1983).

9. W. Jost, *Diffusion in Solids, Liquids, Gases* (Academic Press, New York, 1952).

10. Wikipedia, "Random walk," http://en.wikipedia.org/wiki/Random_walk (accessed May 27, 2006).

11. Y.A. Rylov, "Quantum mechanics as a theory of relativistic Brownian motion," Annalen der Physik **27,** 1 (1971).

12. C. L. Herzenberg, "Quantum Features of Vacuum Flux Impact: An Interpretation of Quantum Phenomena," arXiv.org e-Print archive, physics/0511218 (November 25, 2005), submitted for publication, Physics Essays.

13. K. Namsrai, *Nonlocal Quantum Field Theory and Stochastic Quantum Mechanics* (D. Reidel Publishing Company, Dordrecht Holland, 1986).